\documentclass[aps,amsfonts,prl,showpacs,nobibnotes,nofootinbib,%
tightenlines]{revtex4}
\usepackage{amsmath,amssymb}

\let\cal\mathcal

\newcommand{\beq}{\begin{equation}}
\newcommand{\eeq}{\end{equation}}
\newcommand{\ba}{\begin{array}}
\newcommand{\ea}{\end{array}}

\newcommand{\dd}{Q\!\!\!\!Q}

\begin{document}

\title{Systematics of Exotic Cascade Decays}
\author{Robert Jaffe}
\author{Frank Wilczek}

\affiliation{Center for Theoretical Physics, Laboratory for Nuclear
Science and Department of Physics, Massachusetts Institute of Technology,
Cambridge, Massachusetts 02139\\
MIT-CTP-3463}

\begin{abstract}\noindent Theoretical considerations prompted by
discovery of the exotic $\Theta^{+}(uudd\bar s)$ led us to propose a
dynamical picture emphasizing the role of diquark correlations, which
are also useful in elucidating other aspects of low-energy QCD. A
notable prediction of this picture is the existence of new exotic and
non-exotic $S=-2$ ``cascade'' baryons with specific, characteristic
properties.  We argue here that recent observations by the NA49
collaboration are broadly consistent with our predictions, and propose
further tests.
\end{abstract}

\vspace*{-\bigskipamount}

\maketitle

\section{Introduction}

Recently we proposed that the systematics of exotic baryons in QCD can
be explained by diquark correlations \cite{Jaffe:2003sg}.  If this
picture is correct, the lightest and only prominent $qqqq\bar q$
baryons made of light ($u$,~$d$,~$s$) quarks will form an antidecuplet
of $SU(3)_{f}$ with positive parity, accompanied by a nearly
degenerate octet also containing large $qqqq\bar q$ components.  The
most unusual states in this multiplet, aside from the original
$\Theta^{+}(uudd\bar s)$ \cite{Nakano:2003qx} which motivated the
study, are the quartet of $I(\, {\rm isospin})=3/2$, $S(\,{\rm
strangeness})=-2$, ``cascades'', which we predicted to be quite light
and quite narrow \cite{Jaffe:2003sg}.  The diquark picture is not
the only proposed explanation for the observed exotic baryons.  The
existence of a prominent exotic baryon antidecuplet is a long-standing
prediction of the chiral soliton model\cite{csm}.  Indeed, the
experiment in which the $\Theta^{+}$ was first reported was motivated
by the work of Diakonov, Petrov, and Polyakov \cite{csm}.  Since that
discovery many different models have been proposed for the
$\Theta^{+}$ and related states \cite{many}.

More experimental input and theoretical analysis will be needed to
distinguish among different dynamical pictures of exotic baryons. 
Definitive statements about the internal structure of these new
baryons likely will not be possible until realistic ({\it i.e.\/}
unquenched, light-quark), high-statistics lattice studies are carried
out.  The first lattice studies employed sources which seem to be
poorly matched to the diquark picture \cite{lattice}, but more
appropriate sources have been proposed \cite{jw2} and new studies are
underway.  Below we review the foundations of the diquark
picture and provide a guide to its implications for exotic cascades
and their non-exotic partners.

We also explore, specifically, what can be learned from the decays of
the exotic and non-exotic $qqqq\bar q$ cascade states \cite{rlj}.  Our
work is motivated in part by the recent report of a $\Xi^{--}(ddss\bar
u)$ near 1860 MeV \cite{Alt:2003vb}.  The report needs confirmation. 
On the other hand the $\Theta^{+}$ now seems rather well founded, and given its existence very
general arguments, of which diquark dyanmics are a special case, require light
exotic cascades to till out the antidecuplet.  The phenomenological
implications discussed here follow from the diquark picture of exotic
dynamics and are largely independent of whether or not
Ref.~\cite{Alt:2003vb} is confirmed.  For purposes of concreteness, we henceforth assume that 
the observations reported in Refs.~\cite{Alt:2003vb} and \cite{jlab} 
reflect reality.

Quite a bit can already be inferred from the
observation (and non-observation) of various decay modes reported
by the NA49 collaboration. 
First, of course, the $\Xi^{--}(1860)$ provides further evidence of
the antidecuplet begun with the $\Theta^{+}(1540)$.  Second, the
report of a nearly degenerate $\Xi^{-}(1855)$ decaying into the well
known $\Xi^{*}(1530)$ \emph{and} the apparent absence of a signal for
a nearby $\Xi^{+}$ decaying into the $\Xi^{*}(1530)$\cite{jlab}
together suggest that there is a non-exotic, $I=1/2$, multiplet of
cascades at the nearly the same mass as the $\Xi^{--}(1860)$.  NA49
also reports evidence for a $\Xi^{0}(1860)$ decaying into
$\Xi(1320)\pi$.  While this re-enforces the evidence for narrow
cascades in this mass range, it does not distinguish between $I=1/2$
and $I=3/2$.  The possible existence of an $I=1/2$ multiplet among
this complex of cascades around 1860 MeV could be confirmed by looking
for the decays $\Xi^{-}(1855)\to \Lambda K^{-}$ and $\Xi^{0}(1860)\to
\Lambda \overline K^{0}$, which should be visible in the NA49
apparatus.  When an experiment sensitive to neutral particles
($\pi^{0}$'s and/or neutrons) becomes available, several further
checks will be possible.  Should the existence of both $I=1/2$ and
$I=3/2$ cascade multiplets be confirmed, it would be strong evidence
for the appropriateness of the quark picture of the exotic spectrum,
which requires a roughly degenerate octet and antidecuplet.  By way of
contrast chiral soliton models, while they do, generically, predict the existence
of excited octets, together with many other $SU(3)_{f}$
representations, provide no natural reason for the octet to be
nearly degenerate with the antidecuplet \cite{csm,Diakonov:2003jj}.  Finally, the
observed decay of the $\Xi^{-}(1855)$ provides some indication
concerning its spin and parity.  The spin and parity of the
$\Theta^{+}$ and $\Xi^{--}$ are unknown, although there is some
indication, from the absence of structure in the production angular
distribution \cite{Hicks}, that the $\Theta^{+}$ has $J=1/2$.  A
measurement of the spin and parity would discriminate between
uncorrelated quark models, which predict negative parity, and both
correlated quark models and chiral soliton models, which predict
positive parity.  Specifically, the observation of $\Xi^{-}(1855) \to
\Xi^{*0}(1530)\pi^{-}$ disfavors $J^{\Pi}=1/2^{-}$.
 
\section{Consequences of Diquark Dynamics}

Here we summarize the diquark picture of exotic dynamics and its most
striking predictions.  We assume that quarks, when possible, correlate
strongly in the channel which is antisymmetric in color, spin, and
flavor.  This channel is favored both by gluon exchange
\cite{DeRujula:ge} and by instanton interactions \cite{'tHooft:fv}. 
For light quarks ($u$,~$d$,~and~$s$) the resulting diquark, $\dd$, is
a color and flavor-$SU(3)$ antitriplet with $J^{\Pi}=0^{+}$.  The
correlation is strongest for massless ($u$ and $d$) quarks and
decreases as the mass of one quark or the other increases.  It is
diminished for $ds$ and $su$ pairs.  In the heavy quark limit it
scales like $1/m_{1}m_{2}$, and is probably negligible, except perhaps
for charm-light combinations \cite{Lipkin:1977ie}.  Diquarks with
other color and spin quantum numbers are assumed to be less favored
energetically.   Of course the disfavored diquark (flavor
symmetric, color antisymmetric, $J=1$) also appears in the hadron
spectrum --- most notably in the $3/2^{+}$ baryon decuplet.  A simple
analysis of the masses of strange ($\Lambda$, $\Sigma$, $\Sigma^{*}$)
or charm ($\Lambda_{c}$, $\Sigma_{c}$, $\Sigma^{*}_{c}$) baryons
indicates an approximately 210 MeV energy difference between the
disfavored and the favored ($ud$) diquarks.  This is a significant
difference, enough to make exotic mesons and baryons composed of
disfavored diquarks heavy, broad, and prone to be indistinguishable
from the continuum of ordinary meson and baryon states, into which they
can fall apart without suppression \cite{Jaffe:bu}.  Dominance of the
favored diquark leads to the many predictions for exotic
spectroscopy:

\begin{itemize}
\item No light exotic $qq\bar q\bar q$ mesons will ever be seen,
because the flavor content of $\dd\otimes\overline{\dd}$ is
$3_{f}\otimes \bar 3_{f}=1_{f} \oplus 8_{f}$ 
\cite{Jaffe:1976ig,Jaffe:bu}, which are the same representations as
ordinary $q\overline q$ mesons.  (This does not preclude the
possibility of manifestly exotic meson resonances involving 
favored diquarks with heavy flavors, such as $cs\bar u \bar d$
\cite{Lipkin:1977ie}.)

\item Instead, the only prominent light $qq\bar q\bar q$ mesons will
be a nearly ideally mixed octet and singlet of $J^{\Pi}=0^{+}$~mesons.
These can perhaps be identified with the $f_{0}(600)$, $\kappa(800)$,
$f_{0}(980)$, and $a_{0} (980)$\cite{Jaffe:1976ig}.  These light
scalar mesons have always posed classification problems for quark
models, and there is an entire additional nonet of scalar mesons in
the 1300--1500 MeV range, where $q\overline q$ mesons would be
expected to lie.  Because they are not manifestly exotic however, the
classification of the light scalars as $qq\bar q \bar q$ remains
controversial \cite{smw}.

\item The only light-quark exotic baryons made of four quarks and an
antiquark will lie in an antidecuplet of $SU(3)_{f}$, which will be
nearly-ideally mixed with an octet.  The non-exotic states in these
multiplets will further mix with ordinary $qqq$ baryons.  Since
diquarks are $SU(3)_{f}$ antitriplets, the only way to make an exotic
out of two diquarks and an antiquark is to combine the diquarks
symmetrically in flavor, [$\overline{3}_{f} \otimes
\overline{3}_{f}]_{\cal S} = \overline{6}_{f}$, and then couple the
antiquark.  The flavor content of the resulting $qqqq\overline q$
states is then $\overline 6_{f}\otimes \overline 3_{f}= 8_{f}\oplus
\overline{10}_{f}$ \cite{Jaffe:2003sg}.  Other approaches to exotic
spectroscopy predict a much richer spectrum of exotics including
$27_{f}$ and $35_{f}$ multiplets\cite{many}.  A particularly notable
difference is the absence in the diquark picture of an isovector
analog of the $\Theta^{+}(1540)$, with $S=+1$ and charges $Q=0$, $1$,
and $2$ (a state which occurs in the $27_{f}$ and other exotic
multiplets, but not in the $\overline{10}_{f}$), at low mass, which
seems to be a robust prediction of chiral soliton models \cite{mp} and
which has been sought without success in analyses of $K^{+}p$ data
\cite{plusplus}.

\item The mass splittings of the $[\dd\otimes\dd]_{\cal S}\otimes \bar
q $ octet and antidecuplet baryons, computed to first order in
$m_{s}$, yield a spectrum (as discussed in Ref.~\cite{Jaffe:2003sg})
which includes the $\Theta^{+}(1540)$, two nucleons, $N$ and $N'$, two
$\Sigma$'s, $\Sigma$ and $\Sigma'$, a $\Lambda$ and \emph{two
multiplets of cascades}: one in the antidecuplet with $I=3/2$, which
includes the exotic $\Xi^{+}(uuss\bar d)$ and $\Xi^{--}(ddss\bar u)$,
and the other in the octet with $I=1/2$.  The mass spectrum proposed
in Ref.~\cite{Jaffe:2003sg} follows from the assumption that the
fundamental forces between quarks and antiquarks are flavor
independent.  Then $SU(3)_{f}$ violation introduces one parameter,
$\langle \overline{6}_{f}|\!|{\cal H}_{8}|\!|\overline{6}_{f}\rangle$,
for the $\{qqqq\}_{\overline{6}_{f}}$ and another parameter, $\langle
\overline{3}_{f}|\!|{\cal H}_{8}|\!|\overline{3}_{f}\rangle$ for the
$\bar q$.  The exotic baryon mass does not depend on how the $\bar
6_{f}$ and $\bar 3_{f}$ are finally coupled.  This model gives ideal
mixing: the number of $s+\bar s$ quarks in a hadron is a good quantum
number, so the light nucleon, $N^{+}$ is $uudd\bar d$, while the heavy
nucleon, $N'^{+}$ is $uuds\bar s$.  Likewise the light $\Sigma^{+}$ is
$uuds\bar d$, and the heavy $\Sigma'^{+}$ is $uuss\bar s$.  The masses
of the $8_{f}$ and $\overline{10}_{f}$ baryons are then given by
$M(N)=M_{0}$, $M(\Theta)=M_{0}+\mu$, $M(\Lambda)=M(\Sigma)=M_{0}+\mu
+\alpha$, $M(N')=M_{0}+2\mu+\alpha$,
$M(\Xi_{\overline{10}})=M(\Xi_{8}) = M_{0}+2\mu+2\alpha$, and
$M(\Sigma')=M_{0}+3\mu+2\alpha$, where $M_{0}$ is the common mass in
the $SU(3)_{f}$ symmetry limit, and $\mu$ and $\alpha$ are linear
combinations of the $\bar 3_{f}$ and $\bar 6_{f}$ symmetry breaking
invariant matrix elements.  Ideal mixing is only an approximate
symmetry for well-known mesons ({\it e.g.\/} $\rho$, $\omega$,
$\phi$), so one should not expect high accuracy here.  To emphasize
this we round all masses to the nearest 50 MeV.\footnote{In
Ref.~\cite{Jaffe:2003sg} we took $\alpha=60$ MeV from an analysis of
octet baryon masses, and thereby estimated $M(\Xi)=1750$ MeV.} An
analysis of the complete baryon resonance spectrum \cite{WS} suggests
that the $N(1440)^{1/2^{+}}$ and $\Sigma(1660)^{1/2^{+}}$ should be
identified with the $N$ and $\Sigma$ $qqqq\bar q$ states.  This allows
a determination of $\alpha$ and $\mu$ entirely within the $qqqq\bar q$
sector, with the result $\alpha\approx 100$ MeV and $\mu\approx 100$
MeV. The resulting mass predictions are $M(N)\approx 1450$,
$M(\Theta)\approx 1550$, $M(\Lambda)\approx M(\Sigma)\approx 1650$,
$M(N')\approx 1750$, $M(\Xi_{\overline{10}})\approx M(\Xi_{8})\approx
1850$, and $M(\Sigma') \approx 1950$.  The $\Xi$ mass is closer to the
mass reported by NA49 than our original estimate.  The $N'$ is
predicted at 1750 MeV, close to the $N(1710)^{1/2^{+}}$.  Our model is
obviously crude.  However we know of no framework for multiquark
dynamics -- other than lattice QCD -- which offers a more accurate
analysis.

\item The exotic antidecuplet baryons should have
spin-parity $1/2^{+}$ and be accompanied by nearby states with
$J^{\Pi}=3/2^{+}$ \cite{Jaffe:2003sg,Dudek:2003xd}:  $[\dd\otimes
\dd]_{\cal S}$ must be in the $P$-wave to satisfy Bose statistics.
This $\ell=1$ system can couple to the antiquark to give either
$J^{\Pi}= {3/2}^{+}$ or $ {1/2}^{+}$.

\item Charm and bottom analogues of the $\Theta^{+}(uudd\bar s)$ with
quark content $uudd\bar c$ and $uudd\bar b$ may be stable against
strong decay: The strong decay thresholds for these states depend
on the pseudoscalar meson masses, which grow like the square root
of the quark masses.  Thus, for example, the threshold for
$\Theta^{0}_{c} (uudd\bar c) \to pD^{-}$ is relatively higher than
the threshold for $\Theta^{+}_{s}(uudd\bar s)\to nK^{+}$
\cite{Jaffe:2003sg}.

\item Configurations in which diquarks are in relative $S$-waves will
experience a repulsive interaction due to Pauli blocking
\cite{Jaffe:2003sg}.  States affected by this include the non-exotic
nonet of baryons of the form $[\dd\otimes\dd]_{\cal A} \otimes\bar q$
with negative parity and flavor content $3_{f}\otimes \bar
3_{f}=1_{f}\oplus 8_{f}$, and the $H$-dibaryon, $[\dd\otimes\dd
\otimes\dd]_{\cal A}$, a flavor singlet.  These states will be heavier
and less prominent as a result.
\end{itemize}

Our principal focus here is on the cascade states in the
$SU(3)$-flavor antidecuplet and octet.  We denote the antidecuplet
$I=3/2$ cascade state with charge $Q$ by $\Xi_{3/2}^{Q}$ and the octet
$I=1/2$ cascade state by $\Xi_{1/2}^{Q}$.  The antidecuplet and octet
cascade states share common color and spin wavefunctions and therefore
should be close in mass, except for the possibility that the octet
states could mix with nearby $qqq$ states.  Isospin violating mixing
between the $\Xi_{3/2}^{Q}$ and $\Xi_{1/2}^{Q}$ should be small unless
they are accidentally highly degenerate.  Indeed, the
$\{\Xi^{0}_{3/2}, \Xi^{0}_{1/2}\}$ and $\{\Xi^{-}_{3/2},
\Xi^{-}_{1/2}\}$ are the only pairs of octet and antidecuplet states
with the same charge and strangeness which \emph{should not} mix
significantly.  In contrast, for example, the $N_{\overline{10}}$ and
$N_{8}$ should mix strongly to diagonalize strange quark number.  As a
result all the $qqqq\bar q$ cascade states should respect selection
rules which follow from their isospin and $SU(3)_{f}$ quantum numbers.
The states are expected to have $J^{\Pi}=1/2^{+}$ or
$J^{\Pi}=3/2^{+}$.  If the $\Theta^{+}(1540)$ and the exotic cascades
have the same spin-parity, then their widths can be related by
assuming $SU(3)_{f}$ symmetry for the matrix element and correcting
for phase space.  This should be reliable at the level of typical
$SU(3)_{f}$ symmetry violation, {\it i.e.\/} $\sim 30\%$.   This
estimate gives widths of order 3.5 times the width of the
$\Theta^{+}(1540)$ for exotic cascades with masses of 1860 MeV.
Although no symmetry applies, we also expect the related
$J^{\Pi}=3/2^{+}$ states to be narrow, since the underlying color
dynamics is common to both.

\section{Interpretation of the NA49 Observations}
The cascades observed by NA49 appear to decay into either the
$\Xi(1320)$ ($J^{\Pi}=1/2^{+}$) or the $\Xi^{*}(1530)$
($J^{\Pi}=3/2^{+}$).  For simplicity we refer to the former as the
$\Xi$ and the latter as the $\Xi^{*}$.  To avoid confusion, we will
denote the cascade states observed by NA49 by ``${\mathbf \Xi^{Q}}$''
(where $Q$ is the charge) unprejudiced by theoretical interpretation.
In contrast, when we identify and discuss antidecuplet and octet
states we denote them $\Xi_{3/2}^{Q}$ and $\Xi_{1/2}^{Q}$
respectively.  NA49 can only reconstruct final states without neutral
particles.  Perusal of the PDG tables suggests the following possible
decays of their ${\mathbf \Xi}$ states to $\Xi$ or $\Xi^{*}$ can be
observed:
\begin{alignat}{2}
&\mathbf{\Xi^{--}} &&\to \  \Xi^{-}\ \pi^{-}\label{eq1}\\
&\mathbf{\Xi^{-}} &&\to  \ \Xi^{*0}\ \pi^{-}\label{eq2}\\
&\mathbf{\Xi^{0}} &&\to  \ \Xi^{-}\ \pi^{+}\label{eq3}\\
&\mathbf{\Xi^{+}}  &&\to  \ \Xi^{*0}\  \pi^{+}
\label{eq4}
\end{alignat}
As of this writing, NA49 has presented evidence for decays
(\ref{eq1}), (\ref{eq2}), and (\ref{eq3}).  They have looked for, but
have not seen decay (\ref{eq4}) \cite{ Alt:2003vb,jlab}.  In all cases the
masses are approximately 1860 MeV and the widths are below the
experimental resolution of 18 MeV. Although some of these results are
preliminary and all are unconfirmed, for the purposes of this analysis
we accept them and consider their consequences.  We discuss the flavor
consequences of each decay in turn, and then return to discuss spin and
parity.

\subsection{Flavor Classification}
\subsubsection{ $(1)\ \mathbf{\Xi^{--}}  \to \  \Xi^{-}\ \pi^{-}$}

This decay is allowed by both isospin and $SU(3)_{f}$ symmetry.  It
clearly identifies $\mathbf{\Xi^{--}}$ to be a member of the
antidecuplet with $I_{3}=-3/2$, in our notation $\Xi_{3/2}^{--}$, with
quark content $ddss\bar u$.  $S U(3)_{f}$ symmetry predicts that the
amplitude for $\Xi_{3/2}^{--}\to \Sigma^{-}K^{-}$ is the same as
$\Xi_{3/2}^{--}\to \Xi^{-}\pi^{-}$.  Unfortunately this decay cannot
be seen at NA49 because $\Sigma^{-}\to nK^{-}$, and the neutron cannot
be seen.  The decay $\Xi_{3/2}^{--}\to \Xi^{*-}\pi^{-}$ (which also
cannot be seen by NA49) is forbidden by $SU(3)_{f}$ symmetry because
$10_{f}\otimes 8_{f}\supset\!\!\!\!\!\!/ \ \ \overline{10}_{f} $, and
should be suppressed.  Given that, $\Xi^{-}\pi^{-}$ and
$\Sigma^{-}K^{-}$ are the only two body decay modes for the
$\Xi_{3/2}^{--}$, and its width can be related to the width of the
$\Theta^{+}(1540)$.  At a mass of 1860 MeV, and assuming $P$-wave
phase space, as suggested by the diquark picture, we estimate
   $$
\frac{\Gamma[\Xi^{--}_{3/2}(1860)]}
{{\Gamma}[\Theta^{+}(1540)]} \approx 3.4\quad \mbox{and}\quad
\frac{{\rm BR}[\Xi_{3/2}^{--}(1860) \to \Sigma^{-}K^{-}]}
{{\rm BR}[\Xi_{3/2}^{--}(1860)\to \Xi^{-}\pi^{-}]} \approx 0.5
$$

\subsubsection{ $(2)\  \mathbf{\Xi^{-}} \to  \ \Xi^{*0}\ \pi^{-}$}

This decay is in many ways the most interesting reported by NA49.  It
is tempting to identify the $\mathbf{\Xi^{-}}$ with the
$\Xi_{3/2}^{-}$, the isospin partner of the $\Xi_{3/2}^{--}$.  If so,
the decay to $\Xi^{*0}\pi^{-}$ is allowed by isospin, but the $
\Xi^{-}_{3/2}$ is in the antidecuplet and the decay is forbidden by
$SU(3)_{f}$, since $10_{f}\otimes 8_{f}\supset\!\!\!\!\!/ \ \
\overline{10}_{f} $.  In contrast, the decays $\Xi_{3/2}^{-}\to
\Xi^{0}\pi^{-}/\Xi^{-}\pi^{0}$ (which cannot be seen by NA49) are
allowed by isospin and $SU(3)_{f}$ and predicted to go at the same
rate as $\Xi_{3/2}^{--}\to \Xi^{-}\pi^{-}$.  Furthermore these decays
have more phase space than $\mathbf{\Xi^{-}} \to \ \Xi^{*0}\ \pi^{-}$
(the ratio of $P$-wave phase space factors is approximately 4.5).
Thus if $\mathbf{\Xi^{-}}$ is in the antidecuplet, both phase space and
the $SU(3)_{f}$ selection rule favor $\mathbf{\Xi^{-}}\to
\Xi^{0}\pi^{-}$ over $\mathbf{\Xi^{-}} \to \ \Xi^{*0}\ \pi^{-}$, and
NA49 would have to be seeing a strongly suppressed mode.

Provided the $\mathbf{\Xi^{-}}$ is not produced much more copiously
than $\mathbf{\Xi^{--}}$, and the NA49 sensitivity to the mode
$\mathbf {\Xi^{-}}\to\Xi^{0*}\pi^{-}$ is not much greater than the
sensitivity to $\mathbf{\Xi^{--}}\to\Xi^{-}\pi^{-}$, both of which are
reasonable assumptions, then either $SU(3)_{f}$ is badly violated, or
$\mathbf{\Xi^{-}}$ is not in an antidecuplet.  The first alternative,
$SU(3)_{f}$ violation, predicts a healthy rate for
$\mathbf{\Xi^{+}}\to \Xi^{*0}\pi^{+}$, which has not been seen (see
below).  So we propose that the decay $\mathbf{\Xi^{-}}\to
\Xi^{*0}\pi^{-}$ identifies the $\mathbf{\Xi^{-}}$ to be a member of
an octet, presumably the octet expected in the diquark picture.  If
this is correct, then the NA49 data contain evidence for \emph{both}
octet and antidecuplet cascades near 1860 MeV.

Since this is an important issue, the classification of the $\mathbf
{\Xi^{-}}$ must be confirmed.  Note that the negatively charged
partner, $\Xi^{-}_{3/2}$, of the $\Xi^{--}_{3/2}$ is also expected
to lie in this mass region.  Both the $\Xi_{3/2}^{-}$ and the
$\Xi_{1/2}^{-}$ can decay to $\Xi^{-}\pi^{0}/\Xi^{0}\pi^{-}$, so
this decay channel may show rich structure.  If there were no
$I=1/2$ state then the $\mathbf{\Xi^{--}}$ and $\mathbf{\Xi^{-}}$
are both members of the same $I=3/2$ multiplet and the rate for
$\mathbf{ \Xi^{--}}\to \Xi^{-}\pi^{-}$ and $\mathbf{\Xi^{-}}\to
\Xi^{0}\pi^{-}/\Xi^{-}\pi^{0}$ should be the same, and the ratio of
branching ratios,
$$
\frac{{\rm BR}[\Xi_{3/2}^{-}(1855) \to \Xi^{-}\pi^{0}]}
{{\rm BR}[\Xi_{3/2}^{-}(1855)\to \Xi^{0}\pi^{-}]} = 2,
$$
is determined by isospin symmetry alone. In contrast, if only the
$I=1/2$ octet state $\Xi_{1/2}^{-}$ is present, then the ratio
is inverted,
$$
\frac{{\rm BR}[\Xi_{1/2}^{-}(1855) \to \Xi^{-}\pi^{0}]}
{{\rm BR}[\Xi_{1/2}^{-}(1855)\to \Xi^{0}\pi^{-}]} =\frac{1}{2}.
$$
Deviation from these
simple isospin relations would signal the presence of both $I=1/2$
and $I=3/2$.  Study of this decay channel must await an
experiment sensitive to neutrals.
NA49 can look for the decay
$\Xi^{-}_{1/2}\to \Lambda K^{-}$, which is allowed by $SU(3)_{f}$
and has phase space comparable to the decay $\Xi^{-}_{1/2}\to
\Xi\pi$.  Observation of this decay mode would confirm the
existence of an $I=1/2$ component of the $\mathbf{\Xi^{-}}$,
because $\Lambda K^{-}$ cannot have $I=3/2$.  Unfortunately we cannot
use symmetry to predict a rate for $\Xi^{-}_{1/2}\to \Lambda K^{-}$ on the basis
of the observation of $\Xi^{-}_{1/2}\to \Xi^{*0}\pi^{-}$ or
$\Xi^{--}_{3/2}\to\Xi^{-}\pi^{-}$, because the decays involve
different $SU(3)_{f}$ reduced matrix elements.

\subsubsection{$(3)\ \mathbf{\Xi^{0}}  \to  \ \Xi^{-}\ \pi^{+}$}

Both the $\Xi^{0}_{3/2}$ partner of the $\Xi^{--}_{3/2}$ and the
$\Xi^{0}_{1/2}$ partner of the $\Xi^{-}_{1/2}$ are expected to decay
into $\Xi^{-}\pi^{+}$.  Isospin invariance predicts the rate for
$\Xi^{0}_{3/2}\to \Xi^{-}\pi^{+}$ to be 1/3 that of
$\Xi^{--}\to\Xi^{-} \pi^{-}$.  On the other hand, the rate for
$\Xi^{0}_{1/2}\to \Xi^{-}\pi^{+}$ should be 2/3 of the total rate of
$\Xi_{1/2}^{-}\to \Xi\pi$.  The observation of this decay mode by NA49
confirms the general existence of cascades in the 1860 MeV region, but
does not discriminate between the $I=1/2$ octet and $I=3/2$
antidecuplet.  If we have interpreted decays~(\ref{eq1})
and~(\ref{eq2}) correctly, \emph{both states are needed here}.
Careful measurement and comparisons of the the decays
$\mathbf{\Xi^{-}}\to \Xi^{-}\pi^{+}$ and
$\mathbf{\Xi^{-}}\to\Xi^{0}\pi^{0}$ would help sort out the isospin
structure.  This also awaits an experiment sensitive to neutrals.  The
existence of an $I=1/2$ component in the $\mathbf{\Xi^{0}}$ could be
confirmed by searching for the decay $\mathbf{\Xi^{0}}\to\Lambda
\overline K^{0}$, which could be seen at NA49 as $\Lambda K_{S}$.
Failure to observe it at a level of sensitivity comparable to
decay~(\ref{eq3}) would suggest $I=3/2$.  Unfortunately we cannot use
$SU(3)_{f}$ symmetry to predict the rate for $ \Xi_{1/2}^{0}
\to\Lambda \overline K^{0}$ from the observation of $\Xi_{1/2}^{0}\to
\Xi^{-}\pi^{+}$ because there are two $SU(3)_{f}$ reduced matrix
elements in $8_{f}\otimes 8_{f}\supset 8_{f}$.

\subsubsection{ $(4)\ \mathbf{\Xi^{+}}   \to  \ \Xi^{*0}\  \pi^{+}$}

If the $\mathbf{\Xi^{--}}$ exists, then a $\mathbf{\Xi^{+}}$ must
exist as well.  However decay~(\ref{eq4}) is forbidden by $SU(3)_{f}$
if the $\mathbf{\Xi^{+}}$ is in the antidecuplet.  So its absence is
consistent with the identification of the $\mathbf{\Xi^{--}}$ as the
antidecuplet $\Xi_{3/2}^{--}$.  The decay would not be forbidden if
the $\mathbf{\Xi^{+}}$ were in the $27_{f}$ or $35_{f}$ representation
of $SU(3)_{f}$, so the absence of this decay supports the antidecuplet
assignment of the $\mathbf{\Xi^{--}}$ in contrast to these other possibilities.

Presumably the dominant decays of the $\mathbf{\Xi^{+}}$ are to
$\Xi^{0}\pi^{+}$ and $\Sigma^{+}\overline K^{0}$, neither of which can
be seen at NA49.  Therefore discovery and study of this state must
await an experiment with neutral detection capability.

The absence of a signal for the $\mathbf{\Xi^{+}}$ via
decay~(\ref{eq4}) at NA49 also supports the identification of the
$\mathbf{\Xi^{-}}$ as the octet member $\Xi_{1/2}^{-}$.  The
alternative presented in (\ref{eq2}) above was that the observed decay
$\mathbf{\Xi^{-}}\to\Xi^{*0}\pi^{-}$ is due to $SU(3)_{f}$ symmetry
violation.  But isospin symmetry alone predicts --- on the
antidecuplet hypothesis --- that the rate for
$\mathbf{\Xi^{+}}\to\Xi^{*0}\pi^{+}$ is three times larger than the
rate for $\mathbf{\Xi^{-}}\to\Xi^{*0}\pi^{-}$.  Absence of this signal
at the appropriate level excludes the $SU(3)_{f}$ violation
explanation for decay~(\ref{eq2}).

\subsection{Spin and Parity}

The spin and parity of the $\Theta^{+}(1540)$ and its partners are
unknown and the subject of much speculation \cite{Jaffe:2003sg,csm}.
The decays~(\ref{eq1}--\ref{eq4}), especially decay~(\ref{eq2}), shed
some light on possible spin-parity assignments for the
$\mathbf{\Xi}$'s.  In Table~\ref{tab1} we list the lowest allowed
orbital angular momentum of the two body decay channel as a function
of the initial spin and parity of the $\mathbf{\Xi}$ state up to
$J=3/2$.  If the $\mathbf{\Xi}$'s have positive parity then all the
decays are $P$-waves.  The ratio of $P$-wave phase space for the
decays to $\Xi$ compared to $\Xi^{*}$ is $\approx$ 4.5, which favors
the decays to the $\Xi$ when they are allowed.   For this reason it's
a little surprising that decay~(\ref{eq2}) has been seen.  On the other hand
the relevant invariant matrix elements are unknown, and might well
compensate for the reduced phase space.

\begin{table}
\begin{center}
\begin{tabular}{|r|c|c|c|c|c|}
\hline
$J^{\Pi}$    & $1/2^{+}$& $1/2^{-}$& $3/2^{+}$& $3/2^{-}$&$p_{\rm cm}$\\ \hline
$\mathbf{\Xi^{--}} \to \ \Xi^{-}\  \pi^{-}$&P&S&P&D&445 MeV\\ \hline
$\mathbf{\Xi^{-}}  \to \ \Xi^{*0}\ \pi^{-}$&P&D&P&S&267 MeV\\ \hline
$\mathbf{\Xi^{0}}  \to \ \Xi^{-}\  \pi^{+}$&P&S&P&D&445 MeV\\ \hline
$\mathbf{\Xi^{+}}  \to \ \Xi^{*0}\ \pi^{+}$&P&D&P&S&267 MeV\\ \hline
\end{tabular}
\caption{Orbital angular momentum (in spectroscopic notation) and
center of mass momentum (in Mev) for the $\mathbf{\Xi}$ decay
channels reported by NA49, for various spin-parity assignments of the
$\mathbf{\Xi}$'s.}
\label{tab1}
\end{center}
\end{table}

In contrast, the assignment $J^{\Pi}=1/2^{-}$ for the
$\mathbf{\Xi^{-}}$ seems quite unlikely.  In this case the observed
decay~(\ref{eq2}) would have to be a $D$-wave.  The decays
$\mathbf{\Xi^{-}}\to\Xi^{-}\pi^{0}/\Xi^{0}\pi^{-}$ are $SU(3)_{f}$
allowed, have larger center-of-mass momentum, and are $S$-wave.  So
NA49 would have to have seen a highly suppressed decay of the
$\mathbf{\Xi^{-}}$.  If the observation of decay~(\ref{eq2}) by NA49
stands up, it is evidence against a $1/2^{-}$ assignment for the
$\mathbf{\Xi^{-}}$.  The situation is reversed for $J^{\Pi}=3/2^{-}$:
observation of decay (\ref{eq2}) is more confidently expected
if the $\mathbf{\Xi^{-}}$ has this spin-parity.

To summarize:  the observation of decay (\ref{eq2}) disfavors a $1/2^{-}$
assignment for the $\mathbf{\Xi^{-}}$, is consistent with a $3/2^{-}$
assignment, and makes no strong statement about a $1/2^{+}$ or $3/2^{+}$
assignment.

\section{Summary and Conclusions}

The report of cascade states around 1860 MeV by NA49 has provided
rapid and striking support for the new exotic baryon spectroscopy
initiated by the discovery of the $\Theta^{+}(1540)$ earlier this
year.  The relatively light mass of the $\mathbf{\Xi^{--}}(1860)$
agrees with the prediction of quark dynamics based on diquark
correlations.  Diquark dynamics requires a baryon octet close by the
exotic antidecuplet which includes the $\Theta^{+}(1540)$ and the
$\Xi_{3/2}^{--}(1860)$.  The cascades are a particularly clean system
in which to look for the octet and antidecuplet, because the octet $I=1/2$
and antidecuplet $I=3/2$ cascades are not expected to mix significantly.
Remarkably, the NA49 observation of a $\mathbf{\Xi^{-}}$ decaying to
$\Xi^{*0}\pi^{-}$ and the absence of a $\mathbf{\Xi^{+}}$ decaying to
$\Xi^{*0}\pi^{+}$ suggest the existence of an octet cascade state
nearly degenerate with the $\Xi_{3/2}^{--}(1860)$.  Further experiments can test this assignment
and several other aspects of our theoretical framework.

\subsection{Acknowledgments}
We T. Anticic and K.~Kadija for conversations.
This work is supported in part by the U.S.~Department of Energy
(D.O.E.) under cooperative research agreements~\#DF-FC02-94ER40818.



\end{document}